\documentclass[twocolumn]{aastex63}
\usepackage{newtxtext,newtxmath}

\usepackage{threeparttable}
\usepackage{lineno}
\usepackage{color}

\shorttitle{Luminosity Correlations and $D_{\rm L}(z)$}
\shortauthors{Petrosian, Singal, \& Mutchnick}

\def\beq{\begin{equation}}
\def\eeq{\end{equation}}

\begin{document}

\title{CAN THE DISTANCE-REDSHIFT RELATION BE DETERMINED\\FROM CORRELATIONS BETWEEN LUMINOSITIES?}

\author{V. Petrosian}

\affiliation{Department of Physics and Kavli Institute for Particle Astrophysics and Cosmology, Stanford University\\382 Via Pueblo Mall, Stanford, CA 94305-4060}
\affiliation{Also Department of Applied Physics, Stanford University}

\author{J. Singal}

\affiliation{Physics Department, University of Richmond\\138 UR Drive, Richmond, VA 23173}
\affiliation{Also Visiting Scholar, Kavli Institute for Particle Astrophysics and Cosmology, Stanford University}

\author{S. Mutchnick}

\affiliation{Physics Department, University of Richmond\\138 UR Drive, Richmond, VA 23173}

\email{vahep@stanford.edu $\,$ jsingal@richmond.edu}


\begin{abstract}
We explore whether an independent determination of the distance-redshift relation, and hence cosmological model parameters, can be obtained from the apparent correlations between two different waveband luminosities or fluxes, as has been claimed in recent works using the X-ray and ultraviolet luminosities and fluxes of quasars. We show that such an independent determination is possible only if the correlation between luminosities is obtained independent of the cosmological model, and measured fluxes and redshifts; for example, being based on sound theoretical models or unrelated observations.  In particular, we show that if the correlation is determined empirically for two luminosities obtained from fluxes and redshifts, then the method suffers from circularity.  In case the observed correlation between fluxes in very narrow redshift bins are used as proxy for the luminosity correlation, then we show that one is dealing with a pure tautology with no information on distances and cosmological model.  We argue that the problem arises because of the incomplete treatment of the correlation and we use numerical methods with a joint X-ray and ultraviolet quasar data set to demonstrate this shortcoming. 
\end{abstract}


\section{Introduction} \label{intro}

Recently \citet{RL} -- hereafter RL -- used a measure of the  {\it non-linear correlation} between the X-ray and ultraviolet (UV) luminosities of quasars to arrive at a determination of the shape of the luminosity distance function in the redshift range $1.4<z<5$,  finding that it deviates from that of the $\Lambda$CDM cosmology.  This deviation favored a larger overall matter density fraction $\Omega_m$ and an evolving dark energy equation of state.  Similar results were obtained by the same method in subsequent works using quasar samples at higher redshifts \citep{RL2}, incorporating additional X-ray quasar catalogs \citep{RL3,RLL,RL4}, and joint analyses of quasars along with other cosmological probes \citep{Bargiacci21}. These findings, if true, would be an important new procedure for using extragalactic sources with wide dispersion in their luminosity (i.e.~sources far from being a ``standard candle") for precision cosmological studies on par with near standard candles like Cepheid variables and Type Ia supernovae, but extending to higher redshifts. 

The main aim of this paper is a close scrutiny of the basics of the procedure proposed by RL and to point out some of its shortcomings. Before getting into details it is important to emphasize several crucial aspects of the procedure.

The first is that this is a purely {\it phenomenological} method of using the correlation between {\it observed fluxes}, or deduced {\it apparent luminosities}, that have no direct or obvious relation to the {\it astrophysics} of the sources. That this is true for a correlation between fluxes is obvious.  As has been pointed out in many publications, it is also true for an apparent correlation between luminosities ($L-L$ correlation).  Using non-parametric methods developed by \citet{EP92} (EP) in several publications \citet[e.g.][]{QP2}, \citet[e.g.][]{QP3} and  \citet{V21}, we have shown that the observed $L-L$ correlations (and luminosity and redshift distributions)  quantitatively and qualitatively are very different from the intrinsic ones due to (1) multi-dimensional observational selection effects which truncate the data, (2) the common dependence on the calculated luminosities on distance or redshift, and (3) possible difference in the redshift evolution of the luminosities in different wavebands.
In a recent work \citep{CW}, using analytic methods and simulations, we explored in depth to what extent  apparent correlations in multiwavelength flux-limited data are indicative (or not) of {\it intrinsic} correlations, and hence the physics of  the accretion disks, jets and  characteristics of the super massive black holes, confirming above findings with actual data.  In a more recent paper \citep{PW} we find similar differences between observed and intrinsic $L-L$ correlation in the X-ray and UV wavebands  (the particular wavebands used by RL). An analysis, also using the EP method, by \citet{Dainotti22} agrees qualitatively with this result with somewhat different luminosity evolution rate in the X-ray band.  However, it should be emphasized that the phenomenological method used by RL is independent of such differences between the apparent and intrinsic $L-L$ correlation. As will be clear from the description of the method given below one could use either correlation leading to the same result.

The second is that even in the phenomenological approach, one has to include the effects of the flux truncation (induced by observational selections process) on the observed correlations. This could change the result only quantitatively but not qualitatively. Consequently we will ignore this effect.

Finally, as mentioned above, X-ray/UV  ($L-L$) correlation is not unique. Quasars and other active galactic nuclei (AGNs) show similar non-linear $L-L$ correlations in other pairs of wavebands; radio$-$optical \citep[e.g.][]{QP2}, mid-infrared$-$optical \citep[e.g.][]{QP3} and optical-gamma ray \citep{V21}. Thus, if there is any deviation from $\Lambda$CDM in one pair of wavebands, the same should be true in other pairs.

In this work we explore the question of whether the correlation between two waveband luminosities (or observed fluxes) can be used at all to reliably achieve an independent determination of the distance-redshift relation as was done in RL and subsequent works.  In \S \ref{analytic} we discuss the potential logical issues with such a method.   In \S \ref{DL} we use some numerical analysis to quantify the reasons for the apparent deviation obtained by RL and explore how the procedure can lead to misleading results.  A brief summary and some discussion is presented in \S \ref{disc}. 

\section{Analytic Considerations of The Basic Problem}\label{analytic}

The RL method relies on the assumption that if the degree and form of the correlation between the luminosities in two different wavebands (in their case the X-ray and UV bands) can be deduced, then given observed fluxes in the two bands and redshifts, one can determine the dependence of the luminosity distance function $D_L(z)$ on redshift.  To explore the validity of this method, we use two different possible approaches to the problem.

{\bf A. Starting from $L-L$ correlation:}

Following RL, let us assume that the $L-L$ correlation   can be expressed with a power-law with index $\gamma_L$ and (the dimensionless) proportionality constant $B_L$ as:%
\footnote{We express the luminosities in units of some fiducial luminosity, $L_0$, whose value is irrelevant, but renders the propotionality constant $B_L$ dimensionless.  Note also that RL describe the $L-L$ correlation in log space obtaining $\gamma$ as the slope of the linear regression fit and intercept $\beta_L=\log B_L$.} 

\begin{equation}
L_{\rm x}(z) = B_L \cdot \left( {{ L_{\rm UV}(z) } } \right)^{\gamma_L}.
\label{LLcorr}
\end{equation}  

An important requirement of the procedure proposed by RL is that  $\gamma_L$ and $B_L$ are independent of the cosmological model; they could depend on redshift but not cosmological parameters $\Omega_M, \Omega_\Lambda$ etc. To start with, following RL, we assume that these parameters  are constants (i.e.~independent of also redshift), but as will be evident from the formalism presented below this  assumption is not actually required for the proposed procedure to work.  
The luminoisty-flux relation for a generic waveband $a$ is
\begin{equation}
F_{a} = {L_a K_a(z) \over 4\pi D_L^2(z)} ,
\label{fluxlum}
\end{equation}
where $D_L(z)$ is the luminosity distance and $K_a(z)$ is the K-correction factor.%
\footnote{In some publications \citep[e.g.][]{Bloom2001} the inverse of this is defined as K-correction. We use the original definition given by \citet{OS68} for galaxies.}
To simplify the algebra for the moment we can ignore the small effect of the K-correction, or use rest frame fluxes, $F_r(z)=F(z)/K(z)$, at a well defined frequency band (or monochromatic flux). Then substituting this in Eq. (\ref{LLcorr}) it is easy to show that we obtain the redshift dependence of the luminosity distance as 
\begin{equation}
\label{Ldist}
4\pi D_L(z)^2=\left( {{B_L \, F_{\rm UV}^{\gamma_L}(z) } \over {F_{\rm x}(z) }}\right)^{(1-\gamma_L)^{-1}},
\end{equation}
which, as stressed above, would be the case even when the fit parameters $B_L$ and $\gamma_L$ depend on redshift {\it but not the cosmological model}.  RL then used the measured fluxes in X-ray and UV bands to determine the luminosity distance as function of redshift.%
    \footnote{It is clear that the method fails for {\it a linear correlation}, $\gamma_L=1$, thus the requirement of nonlinearity. In what follows we assume that $\gamma_L < 1$, which seems to be the case of interest here,  using $L_X$ as the dependent variable. In the opposite case one would have $\gamma_L >1$.}

However, there are many unsupported assumptions in this procedure, the most important of which is that the correlation form in equation (\ref{LLcorr}) {\it is independent of a cosmological model}. In general, the $L-L$ correlations are usually based on luminosities determined from observed (or rest frame) fluxes using Equation (\ref{fluxlum}), which requires an assumed cosmological model that gives the required luminosity distance function $D_L(z)$ (as was done to plot the luminosity vs redshift in Figure 1 of RL).  In that case it is obvious that the above procedure is {\it logically circular} and the method should return the assumed form of the luminosity distance used in calculating the luminosities, modulo observational uncertainties, numerical errors, and neglect of truncation effects.

If one uses the luminosity evolution (LE) corrected luminosities $L'_a=L_a/g_a(z)$, where $g_a(z)$ describes the LE in waveband band $a$, one could follow the above steps using the intrinsic (or de-evolved) $L^\prime-L^\prime$ correlation, with $B'(z)$ and $\gamma'(z)$ possibly not equal to $B(z)$ and $\gamma(z)$.  This leads to the same final results (Eq. \ref{Ldist}) with the addition of LE factors of $g_{\rm x}(z)$ and $g_{\rm UV}(z)$, but still suffering from the same circularity because one must assume a form for $D_L(z)$ to determine the LE functions and the intrinsic correlation $\gamma'_L$. \citet{Dainotti22} use this (circular) procedure and obtain similar result as RL. This is probably because the LE factors they obtain for X-ray and UV bands are similar, which yields $\gamma'(z)\sim \gamma(z)$. See also discussion below.

{\bf B. Starting from $F-F$ correlation:}

\begin{figure}
\hspace*{-0.1in} 
\includegraphics[width=3.5in]{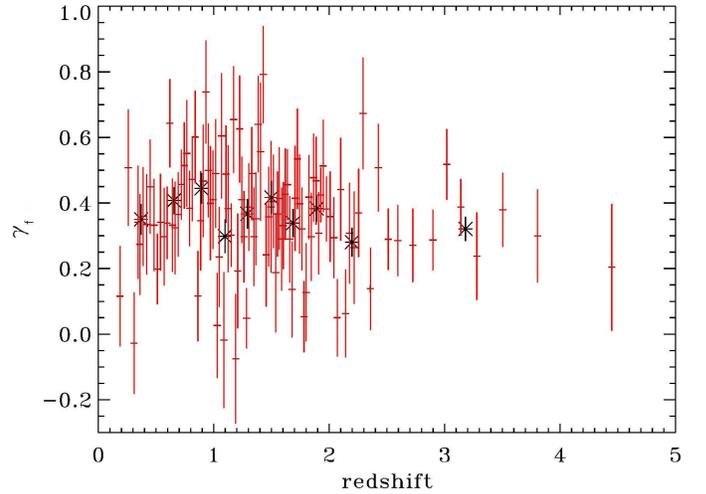}
\caption{A determination of the best-fit power law correlation between the X-ray and UV {\bf fluxes} of quasars, of the form of equation (\ref{FFcorr}), as determined for the data set assembled in \citet{PW}.  Results, with statistical uncertainties, are shown for both ten (stars) and 100 (red pluses) bins of redshift with equal numbers of objects, in respective bins.  The average value in each case is close to $\gamma_{f}$=0.36.  These results evidence significantly larger dispersion that those obtained by RL.}
\label{gammaf}
\end{figure}

On page 2 of their paper RL indicate that their calculation of the correlation index $\gamma$ is ``cosmologically independent'' -- i.e.~independent of a specific cosmological model and therefore of the luminosity distance -- which must mean that in their figure showing $\gamma$ vs redshift (supplementary Figure 16) the correlation indicies are obtained using the correlation between {\it fluxes}, $F_{\rm UV}$ and $F_{\rm x}$, in redshift bins $z_i$ to $z_i+\Delta z_i$  
assuming a power-law correlation
\begin{equation}
F_{\rm x} = B_f(z) \cdot \left( { F_{\rm UV} }  \right)^{\gamma_f(z)}.
\label{FFcorr}
\end{equation}
In general, both the proportionality constant $B_f$ and index, $\gamma_f$ could vary with redshift. In Figure 8 of the supplementary material RL show variation of both quantities with redshift. There are small variations of $\gamma_f$, with an average value of $\gamma_f=0.607 \pm 0.05$, and $\beta_f(z)=\log B_f(z)$ increasing with redshift almost monotonically from 3.8 to 4.5 over the redshift range $z=0-4$.%
\footnote{There is some apparent ambiguity as to whether RL use a value of $\gamma=0.633$ as seems to be indicated in their main paper or $\gamma=0.607$ as indicated in their supplementary paper, but the precise value is not important for our analysis.}
RL use a constant value for $\gamma_f$, but it is not clear if  same is assumed for $B_f(z)$, or whether its variation with redshift is included in the analysis. As will be clear, the exact values of these parameters and whether they are constant or vary with redshift, does not change the arguments presented below. In fact, using a larger sample of quasars analyzed in \citet{PW}, we find large variation of $\gamma_f(z)$ (especially for small redshift bins) with a smaller mean of 0.36 shown in Figure \ref{gammaf}. 

Starting from Equation (\ref{FFcorr}), it is important to note that correlation parameters calculated from fluxes cannot be substituted for correlation parameters needed for luminosities, because of the $L-F$ relation depends on $D^2_L(z)$. 

Let us first consider the relation between the two power-law indexes, $\gamma_f(z)$ and $\gamma_L(z)$ (which for maximal generality we will let vary with redshift), and the possibility that  $\gamma_L(z)=\gamma_f(z)$ independently from the cosmological model.  
The first possibility  is  if  $\gamma_f$'s are are obtained using infinitesimally small redshift bins, within which variation of $D^2_L(z)$ is small and can be ignored.  From Figure 16 in their supplementary paper it appears that RL are using ten bins spanning the redshifts from 0 to 4 with the largest bin  spanning from $z=\sim 2.7$ to $z=4$. The above assumption is clearly not valid for this bin with $\Delta D^2_L/\langle D^2_L\rangle\sim 0.8$ (and not negligible), where the claimed deviation of the luminosity distance model from $\Lambda$CDM is most significant. 

The second possibility is that we have  a prior knowledge that the LE in the two bands are {\it identical}.  In that case the $L-L$ correlation will be a scaled (by $4\pi D_L^2(z)$) version of the $F-F$ correlation yielding $\gamma_L(z)=\gamma_f(z)$. We note that RL and many subsequent papers ignore the possibility of LE. As mentioned above \citet{Dainotti22} find somewhat similar (but not identical) LE for X-ray and opt-UV bands, indicating that this equality is approximately true. However,  in \citet{PW}, using a more rigorous accounting of the X-ray flux threshold we obtain very different LEs.   

But even assuming that these uncertainties can be ignored, let us follow RL and set $\gamma_L(z)=\gamma_f(z)=\gamma(z)$, assuming that $\gamma_L(z)$ is independent of cosmological model, and consider the relation between the two proportionality constants. If we replace the fluxes in equation (\ref{FFcorr}) by luminosities using equation (\ref{fluxlum}), still ignoring the K-corrections, we obtain
\begin{equation}
L_x(z)=B_f(z) \, \times L_{\rm UV}^{\gamma(z)}(z) \, \times [4\pi D^2_L(z)]^{1-\gamma(z)}.
\label{NewLLcorr}
\end{equation}
Comparing this with the required relation in Equation (\ref{LLcorr}) we find
\begin{equation}
B_L(z)=B_f(z) \, \times [4\pi D^2_L(z)]^{1-\gamma(z)},
\label{Blz}
\end{equation}
which clearly is not independent of the luminosity distance and the cosmological model, a strict requirement of the procedure to yield an independent determination of the luminosity distance function. If now one substitutes this form of $B_L(z)$ in equation (\ref{Ldist}) the luminosity distance $D^2_L(z)$ simply cancels out and one recovers equation (\ref{FFcorr}), which is where we started from. Thus, we conclude that  we are dealing with {\it pure tautology}, and that there is no logical procedure for recovering $D_L(z)$.  We emphasize again that this conclusion is true whether or not $\gamma(z)$ varies with redshift or is actually a constant, and even if one has infinitesimally small  redshift bins and identical LE in the two bands so that the assumption $\gamma_L(z)=\gamma_f(z)=\gamma(z)$ is valid. 

{\bf In summary} the procedure proposed by RL is either logically circular or pure tautology,  leading to the inevitable conclusion that it is not possible to arrive at an independent determination of the luminosity distance function $D_L(z)$ using either the correlation index $\gamma_L$ calculated from the luminosities or $\gamma_{f}$ calculated from the fluxes.  The former has an inescapable a-priori dependence on the form of $D_L(z)$ resulting in a logical circularity, and the latter has no relation to $D_L(z)$.

One may ask why, in light of the circular or tautological aspects discussed above, did RL not obtain the $D_L(z)$ function of the $\Lambda$CDM cosmology (or fail numerically) when they carried out their analysis.  We explore this issue in the next section.

\section{Exploring the RL Method With Quasar Data} \label{DL}

Given the considerations of \S \ref{analytic}, any analysis that attempts to use flux-redshift data to determine $D_L(z)$ based on equation (\ref{Ldist}), should return the luminosity distance function of the assumed cosmological model (here $\Lambda$CDM).  However, RL and subsequent works (e.g.~ \citet{Dainotti22}, \citet{KR20}, \citet{KR20}) obtained a $D_L(z)$ function which deviates from that of the $\Lambda$CDM cosmology, favoring a larger value for the matter density $\Omega_m$ and an evolving dark energy equation of state.

The formalism presented above indicates that there is a degeneracy between the assumed redshift dependencies of $D_L$ and the correlation parameters $B$ and $\gamma$, irrespective of whether we start with equation (\ref{LLcorr}) or (\ref{FFcorr}). It is clear from equations (\ref{Ldist}) and (\ref{NewLLcorr}) [or (\ref{Blz})] that this degeneracy takes the form of 
\begin{equation}
\label{DzBzRelation}
D_L(z)\propto B_f(z)^{{1}\over {2(\gamma(z)-1)}}.
\end{equation}
This indicates that an important source of this discrepancy found by RL is that they did not include the redshift dependence of $B_f(z)$. 

As the above equation shows, an increase  with $z$ of $B_f(z)$, and/or decrease of $\gamma(z)$ (for $\gamma(z)<1$), can give rise to a value for $D_L(z)$ that is increasingly lower at higher $z$'s than the true luminosity distance. In the opposite case one would get a higher  $D_L(z)$. For example, an increase of $\beta(z)=\log B_f(z)$ from 3.8 to 4.5 $(\sim 20\%$) obtained by RL, mentioned above, will cause a $\sim 25\%$ decrease in $D_L$ for $\gamma\sim 0.6$, which is about what RL obtained.

\begin{figure}
\includegraphics[width=3.5in]{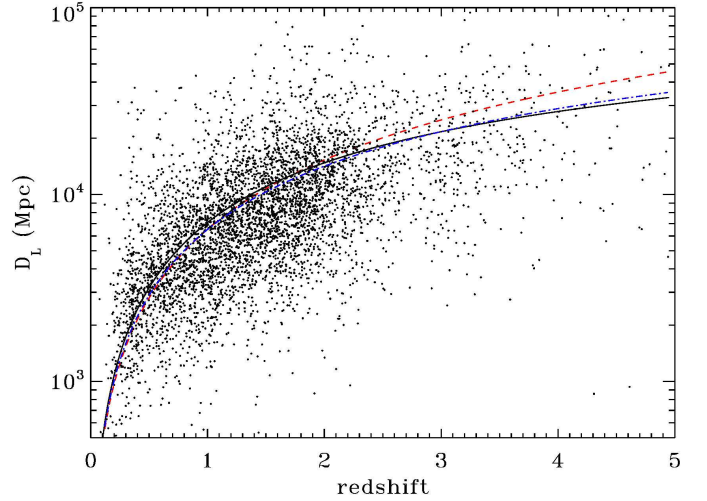}
\caption{$D_L(z)$ as would be determined by equation (\ref{Ldist}), the method of RL, utilizing an X-ray and UV data set of quasars analyzed in \citet{PW}.  The points show $D_L(z)$ for individual quasars as determined for $\gamma=0.28$ obtained by \citet{PW}, while the solid black line shows the best-fit curve of the form equation (\ref{DLfiteq}) for these points.  The dash-dot blue line is the result reported by RL for $\gamma=0.633$.  The dashed red line shows $D_L(z)$ for $\Lambda$CDM.  As evident a lower value of $\gamma$ yields a larger deviation from $\Lambda$CDM at high $z$, as discussed in \S \ref{DL}. }
\label{bestDL}
\end{figure}

To further demonstrate this effect, in Figure \ref{bestDL} we compare the luminosity distance as would be obtained by equation (\ref{Ldist}) for two values of the index $\gamma$:  $\gamma=0.633$ obtained by RL and $\gamma=0.28$ that we obtain using the intrinsic $L'-L'$ correlation obtained in \citet{PW}.  For this purpose, following RL, we assume no redshift dependence for both $B_f$ and $\gamma$.   As in RL, we fit the determined $D_L(z)$ for $\gamma=0.28$ points to a third order polynomial in $\ln(1+z)$, fixing the terms so that it approaches $D_L(z)=(c/H_0)z$ for $z\ll 1$:  
\begin{equation}
D_L(z) =\ln (10) \, (c/H_0)(x+a_2x^2+a_3x^3), \,\,\,\,{\rm with} \,\,\,\, x=\log(1+z).
\label{DLfiteq}
\end{equation} 

Figure \ref{bestDL} shows the result of this fit as the black curve.  We also show  $D_L(z)$ curve obtained by RL for $\gamma=0.63$ (dash-dot blue line) and for $\Lambda$CDM (red-dashed). A larger deviation from $\Lambda$CDM is obtained (at high $z$) for a smaller value of $\gamma$ (for $B(z)$ increasing with $z$) as predicted by equation (\ref{DzBzRelation}). An additional possible source of the apparent deviation in $D_L(z)$ from that of $\Lambda$CDM, investigated by \citet{Yang20} and \citet{Banerjee21}, is that the polynomial expansion of equation (\ref{DLfiteq}) generically fails to recover flat $\Lambda$CDM beyond $z \sim 2$.  Furthermore, \citet{KR20} have shown that the deviation in $D_L(z)$ from that of $\Lambda$CDM in RL is not as statistically significant as claimed.

\section{Summary and Discussion} \label{disc}

We find in \S \ref{analytic} that the method of determining the luminosity distance as a function of redshift (and hence cosmological parameters)  from the observed (non-linear) correlation between two luminosities  or fluxes in a population quasars, proposed by RL and utilized in the subsequent works \citet{RL2}, \citet{RL3}, and \citet{RL4}, suffers from a fatal logical inconsistency inherent in the method.  It is either circular or tautological. 

Determinations of the luminosity distance function commonly involve establishing a correlation between i) a distance independent characteristic, such as the pulsation periods of Cepheid variable stars or the decay times of Type-Ia supernovae, and ii) a distance-dependent quantity, such as the luminosity in a waveband (which is distance-dependent via the flux-luminosity relation).  If such correlations are  redshift, or in general distance, independent, then one can use the ``Hubble diagram'' to get the luminosity distance, otherwise one needs to take into account how the relevant parameters may change with redshift. This condition is more difficult to establish in attempts to use gamma-ray bursts utilizing relationships between the peak spectral energy and the total energy or peak luminosity  \citep[e.g.][]{Amati08,Yon04}.  The method of \citet{RL} and subsequent works is significantly different as it attempts to use the non-linear correlation between two distance-dependent quantities, namely luminosities in two different wavebands (here X-ray and UV).  If valid this would be a new revolutionary method of determining cosmological parameters from extragalactic sources, such as quasars, or more general AGNs, with rich observations at many different wavebands from radio to gamma rays.

We have argued that the only way the proposed method would work is if the exact form and parameters of correlation  between two luminosities is determined {\it independent of the cosmological model} (and hence $D_L(z)$), {\it and without use of the observed fluxes}, such as would be obtained from a purely theoretical physical model of the emitting system.  If the $L-L$ correlation is obtained, as it is commonly the case, from measured fluxes and redshifts, which requires a knowledge of $D_L(z)$, then the method suffers from circularity and should return the assumed $D_L(z)$, modulo observational and numerical errors, and other uncertainties.   An alternative approach of \textit{simultaneous} optimization over \textit{both} cosmological parameters and luminosity correlation parameters as was done by \citet{KR21} and \citet{KR22} is a promising possibility, which apparently leads to no statistically significant deviation from $\Lambda$CDM.

However, RL  obtain correlation parameters using redshift-binned flux-flux correlations. They assume a power law correlation with two parameters; the power law index $\gamma_f(z)$ and proportionality parameter $B_f(z)$ in equation (\ref{FFcorr}), both of which show some variation with redshift in their analysis.  The basic idea they propose is that given infinitesimally small redshift bins, within which the $D_L(z)$ change is negligible, then one can use these parameters for the same form of the $L-L$ correlation with $\gamma_L(z)=\gamma_f(z)$ and $B_L(z)=B_f(z)$.  As we discuss in \S \ref{analytic}, there are several problems with this procedure:

\begin{enumerate}

\item Some redshift bins used by RL span a wide range of redshift and cannot be consider infinitesimal. As shown in Figure \ref{gammaf} using smaller bins one obtains much larger variation of $\gamma_f$ than the ten bins used by RL.  Additionally we obtain a signficantly different value for $\gamma_f$ than RL in our analysis with a joint X-ray and UV dataset of quasars. 

\item RL ignore the small variation of $\gamma_f$ shown by their data. Nevertheless let us follow RL and ignore this variation and set $\gamma_L=\gamma_f=\gamma$.

\item They also seem to neglect the relatively significant monotonic variation they obtain for $B_f(z)$, and, most importantly, assume that the same is true for $B_L(z)$. As we have shown this is manifestly not well founded and when changing variables from fluxes to luminosities (related through $D_L(z)$) one gets a proportionality constant $B_L(z)$ that depends on $D_L(z)$ (independent of whether variation of $B_f(z)$ is significant or not) in a way that $D_L(z)$ drops out of the relation making the argument pure tautology.  

\end{enumerate}

Finally, in \S \ref{DL} we show that the likely reason RL obtained a luminosity distance function which deviates from the $\Lambda$CDM model is related to the fact that they  ignore the redshift dependence of $B_f(z)$ and additionally, because of a potentially questionable method of fitting of $D_L$ to redshift.  As we demonstrate the deduced $D_L(z)$ depends on $B_f(z)$ and $\gamma(z)$ (see, eq.\ref{DzBzRelation}) in a way that explains the deviation downward from the $\Lambda$CDM model that they find at high $z$, shown in Figure \ref{bestDL}.  It is our conclusion then that procedure proposed by RL is not an appropriate method for constraining the luminosity distance function or cosmological parameters.

\acknowledgments

This work relies in part on data analysed in \citet{PW} which was obtained from the Sloan Digital Sky Survey (SDSS), and the Chandra and XMM-Newton X-ray observatories.  Funding for the SDSS and SDSS-II has been provided by the Alfred P. Sloan Foundation, the Participating Institutions, the National Science Foundation, the U.S. Department of Energy, the National Aeronautics and Space Administration, the Japanese Monbukagakusho, the Max Planck Society, and the Higher Education Funding Council for England. The SDSS Web Site is http://www.sdss.org/.  This research has made use of data obtained from the Chandra Source Catalog, provided by the Chandra X-ray Center (CXC) as part of the Chandra Data Archive.  This research has made use of data obtained from the 4XMM XMM-Newton Serendipitous Source Catalog compiled by the 10 institutes of the XMM-Newton Survey Science Centre selected by ESA.

\end{document}